\documentclass{sigchi-ext}
\usepackage[T1]{fontenc}
\usepackage{textcomp}
\usepackage[scaled=.92]{helvet} % for proper fonts
\usepackage{graphicx} % for EPS use the graphics package instead
\usepackage{balance}  % for useful for balancing the last columns
\usepackage{booktabs} % for pretty table rules
\usepackage{ccicons}  % for Creative Commons citation icons
\usepackage{ragged2e} % for tighter hyphenation
\usepackage{caption}

\def\plaintitle{Visualizing the Weird and the Eerie} 
\def\emptyauthor{}
\def\plainkeywords{Data visualization; reflection; affect; agency; semantics.}

\title{Visualizing the Weird and the Eerie}

\numberofauthors{1}
\author{%
  \alignauthor{%
    \textbf{Matthew Brehmer}\\
    \affaddr{Tableau Research}\\
    \email{mbrehmer@tableau.com} }
}

\definecolor{linkColor}{RGB}{6,125,233}
\hypersetup{%
  pdftitle={\plaintitle},
  pdfauthor={\emptyauthor},
  pdfkeywords={\plainkeywords},
  bookmarksnumbered,
  pdfstartview={FitH},
  colorlinks,
  citecolor=black,
  filecolor=black,
  linkcolor=black,
  urlcolor=linkColor,
  breaklinks=true,
}

\begin{document}

%% For the camera ready, use the commands provided by the ACM in the Permission Release Form.
% \CopyrightYear{2023} 
% \setcopyright{rightsretained} 
% \conferenceinfo{alt.VIS}{October 2023, Melbourne, Victoria, Australia}
% \copyrightinfo{\acmcopyright}

\CopyrightYear{2023}
\setcopyright{rightsretained}
\conferenceinfo{alt.VIS}{October 2023, Melbourne, Victoria, Australia}
\isbn{}
% \doi{}
%% Then override the default copyright message with the \acmcopyright command.
\copyrightinfo{\acmcopyright}

\maketitle

% Uncomment to disable hyphenation (not recommended)
% https://twitter.com/anjirokhan/status/546046683331973120
\RaggedRight{} 

% Do not change the page size or page settings.
\begin{abstract}
  In this brief essay, I reflect on how Mark Fisher's definitions of the \textsl{weird} and the \textsl{eerie} could be applied in communicative data visualization. I ask how visualization designers might elicit these two impressions when a viewer is engaging with multimodal representations of data. I argue that there are situations in which viewers should feel uncertain or suspicious of unseen forces that account for the presence or absence of audiovisual patterns. Finally, I conclude that the ability to appreciate the weird and the eerie in data is particularly important at this moment in history, one marked by significant ecological and economic disruption.
\end{abstract}

\keywords{\plainkeywords}

\section{What is Weird? What is Eerie?}

In his 2016 book \textsl{The Weird and the Eerie}~\cite{fisher2016}, the late writer Mark Fisher set out to define these two constructs as both affects and as modes: \textsl{``modes of film and fiction, modes of perception, ultimately, you might even say, modes of being.''}
In this collection of essays, he references a selection of critically notable novels, films, television series, and pieces of music that evoke these two affects.
His writing led to my own realization that data visualization may also be capable of instilling a sense of the weird and the eerie.

Fisher argues that the weird and the eerie are distinct from one another, even though they are both often associated with (and conflated in) the genres of horror, science fiction, and especially post-apocalyptic speculative fiction.
However, the weird and the eerie are not necessarily horrific or devastating. 

% \marginpar{\vspace{-6pc}\textbf{Author's note}: I refrain from using images from visualization projects in this essay, as weird and eerie aspects of visualization are often situated, dynamic, or interactive and cannot be easily captured by still images. Instead, I use thematically suggestive images that have permissive licenses (e.g., CC BY-SA or public domain).\\}

% \begin{marginfigure}
%   \begin{minipage}{\marginparwidth}
%     \centering
%     \vspace{1pc}
%     \includegraphics[width=\marginparwidth]{figures/sigmund-HMCNkAK45r0-unsplash.jpg}
%     \captionsetup{labelformat=empty}
%     \caption{\textbf{Credit}: Sigmund (Unsplash).}~\label{fig:marginfig}
%   \end{minipage}
% \end{marginfigure}

According to Fisher, the \textbf{weird} is \textsl{``that which does not belong,''} that is, the presence of something that seems \textsl{wrong} or at least \textsl{strange}. 
In this sense, the weird can be an object of fascination, something to be sought out and enjoyed. 
I think of my own enjoyment of certain music or novels that exhibit this sense of strangeness. 
Similarly, young children are particularly amused by games of \textsl{``one of these things is not like the other.''} 
But the weird goes beyond the placement of unfamiliar things within a familiar context; the weird is also invoked when familiar things appear out of place, or when a familiar ordering of events is seemingly violated.
To the extent that events can be weird, Fisher remarks that an archaic meaning of weird refers to \textbf{fate}, or a \textit{``twisted form of time and causality.''}

% \begin{marginfigure}
%   \begin{minipage}{\marginparwidth}
%     \centering
%     \includegraphics[width=\marginparwidth]{figures/krzysztof-kotkowicz-OKPhtQsyvM8-unsplash.jpg}
%     \captionsetup{labelformat=empty}
%     \caption{\textbf{Credit}: K. Kotkowicz (Unsplash)}~\label{fig:marginfig}
%   \end{minipage}
% \end{marginfigure}

In contrast to the weird, Fisher defines the \textbf{eerie} as a reaction to situations \textsl{``when there is something present when there should be nothing, or when there is nothing present when there should be something.''} 
In either manifestation of the eerie, there arises a question of \textbf{agency}, a question of who or what is responsible for the lack of absence or the lack of presence, and ultimately a question of what motivated --- or continues to motivate --- this unseen agent.

\section{Visualization Design, Affect, and Semantics}

This essay is motivated in part by a renewed discussion of affect in visualization design.
There are undoubtedly works of data visualization that elicit positive affective responses; examples include the playful and humorous \textsl{xkcd} comic by Randall Munroe~\cite{xkcd}, Tyler Vigen's \textsl{Spurious Correlations}~\cite{vigen2015}, or the oeuvre of Nigel Holmes~\cite{holmes2023}.
Meanwhile, other presentations about data can evoke profound sadness, such as in Neil Halloran's multimodal data documentaries~\cite{Halloran}.
Beyond the polarities of joy and sadness and simple models of affect spanning the dimensions of valence and arousal, thoughtful visualization design can elicit a wider gamut of affective responses, as noted in a recent paper by Elsie Lee-Robbins and Eytan Adar~\cite{lee2022affective}. 
The most concrete affective learning goal in communicative visualization is persuading viewers to take action~\cite{pandey2014persuasive}, or change behavior based on an affective response to a representation of data. 
To compel to action, however, first requires strengthening or changing viewers' beliefs. 
Encountering the weird is an opportunity to change what one believes; Fisher writes that the very existence of the weird object results in the conclusion that \textsl{``the weird thing is not wrong} [---] \textsl{it is our conceptions that must be inadequate.''} 

Another motivation for this essay stems from my interest in visualization techniques that integrate semantic cues into encodings of data.
For instance, the cardinal direction `up' might be good in one chart (e.g.,~profits are rising), whereas up is bad in another chart (e.g.,~global average temperatures are rising); annotations, captions, color palettes, iconography, and other graphical elements can certainly reinforce these good / bad semantics.
The use of figurative elements and frames~\cite{byrne2019figurative} can also serve as contextual cues, suggesting the topic of the visualization upon a cursory glance.

Beyond suggesting semantic associations or that particular visual patterns exhibited by the data are good or bad, it is less clear how to effectively communicate that some patterns are simply \textsl{weird} or strange.
Similarly, it is not universally obvious how visualization designers can communicate that a particular pattern is present or absent, as well as how to express a lack of an explanation for these presences or absences. 
The impression of weirdness or eeriness will dissipate if an explanation is offered, such as when we annotate erratic spikes and troughs in line charts with contextual information external to the data~\cite{hullman2013contextifier}.

\section{Related Affects in Experience Design}

% \begin{marginfigure}
%   \begin{minipage}{\marginparwidth}
%     \centering
%     \includegraphics[width=\marginparwidth]{figures/christopher-ott-7VNWxnaC2hk-unsplash.jpg}
%     \captionsetup{labelformat=empty}
%     \caption{\textbf{Credit}: C. Ott (Unsplash)}~\label{fig:marginfig}
%   \end{minipage}
% \end{marginfigure}

% \begin{marginfigure}
%   \begin{minipage}{\marginparwidth}
%     \centering
%     \includegraphics[width=\marginparwidth]{figures/anderson-schmig-R4zbB0p8-v0-unsplash.jpg}
%     \captionsetup{labelformat=empty}
%     \caption{\textbf{Credit}: A. Schmig (Unsplash)}~\label{fig:marginfig}
%   \end{minipage}
% \end{marginfigure}

Before I reflect further on the weird and the eerie in the context of visualization design, I'll briefly consider related affects from a broader human-computer interaction perspective.
One particularly memorable alt.CHI session that I attended in 2011 included Steve Benford and Brendan Walker's talk about the design of \textsl{fearsome} interactive experiences~\cite{marshall2011gas}, i.e., those that instil a sense of terror that could be implemented in environments such as virtual reality games, escape rooms, or amusement parks. 
As mentioned earlier, the weird and the eerie ere not necessarily terrifying, but I expect that some of the visual and interaction design patterns that elicit terror could be applied when visualizing the weird and the eerie. 
Consider narrative presentations about data, such as data videos or scrollytelling articles: the pacing or delivery could establish a sense of tension, delaying or withholding any revelatory explanation, prompting viewers to imagine or anticipate the forces responsible for weird and eerie audiovisual patterns.
An egocentric perspective could accentuate this tension; for an example of such a perspective, I encourage the reader to \textit{Ride the Nasdaq}~\cite{KennyBecker} (in which the \textit{The Wall Street Journal}'s Roger Kenny and Ana Asnes Becker transform a line chart into a 3D roller coaster ride).

Another related affect, one less visceral than terror, is that of a sense of uncertainty induced by \textbf{ambiguity}. 
William Gaver and colleagues~\cite{gaver2003ambiguity} proposed a set of guidelines for eliciting this reaction to an ambiguous interactive experience, and of these, two are particularly applicable to eliciting a sense of the weird or the eerie in data. 
These guidelines are \textit{``point}[ing] \textit{out things without explaining why''} and \textit{``cast}[ing] \textit{doubt on sources to provoke independent assessment.''}
In observing these guidelines, attention can be drawn to the weird without providing the comfort of an explanation.
Similarly, a visualization designer can elicit alternate hypotheses for a failure of absence or a failure of presence; of course, unintentional human error leading to inconsistent data collection could be to blame, but there could also an unseen agent at play, or an agent that is compromising the processes of data collection and retention.

\section{Visualizing the Weird}
Visualization design consultant Andy Kirk remarked~\cite{Kirk2014} that people \textsl{``are naturally drawn to gaps and exceptions and things that don't really fit in with the rest.''} 
And yet, designing representations of data that emphasize these aspects can be \textsl{``a very fiddly challenge.''}

In a paper about scatterplot matrices, Leland Wilkinson and colleagues~\cite{wilkinson2006high} observed that real datasets \textsl{``frequently contain} [\ldots] \textsl{outliers, missing data, and ``\textbf{just plain weird}'' bivariate distributions''} [emphasis added].
A professional data analyst is trained to spot these weird distributions, distributions that seem strange or wrong given the context of the data. 
But how can the analyst effectively communicate that these distributions are weird to a lay audience? 

I echo the argument of Enrico Bertini and colleagues~\cite{bertini2020shouldn}, that despite the precision and generalizability that scatterplots and their variations offer, using other forms of representation can promote \textsl{``serendipitous discovery, educational impact, \textbf{hedonic response}, or changes in behavior''} [emphasis added]. 
As with the hedonic enjoyment of weird music or novels, I similarly enjoy encountering \textsl{weird} representations of data. 
I admit that part of this enjoyment can be attributed to what Fisher describes as \textsl{``seeing the familiar and the conventional becoming outmoded,''} which in my case is a rejection of dogmatic minimalism and the prioritization of perceptual precision in visualization design. 
Beyond this enjoyment is an appreciation that some weird choices of representation can accentuate weird patterns in data. 
For instance, consider the connected scatterplot~\cite{haroz2015connected}: it can resemble a line chart, but it can also exhibit visually prominent line reversals and loops, indicative of weird events in the relationship between two variables. 
Absent weird events, the connected scatterplot is often a poor choice for showing bivariate temporal relationships~\cite{kosara2016presentation}.

% \begin{marginfigure}
%   \begin{minipage}{\marginparwidth}
%     \centering
%     \includegraphics[width=\marginparwidth]{figures/Grotesqueengraving.jpg}
%     \captionsetup{labelformat=empty}
%     \caption{\textbf{Credit}: V\&A Museum (Wikimedia)}~\label{fig:marginfig}
%   \end{minipage}
% \end{marginfigure}

% \begin{marginfigure}
%   \begin{minipage}{\marginparwidth}
%     \centering
%     \includegraphics[width=\marginparwidth]{figures/manikandan-annamalai-RvSbLuRqiGw-unsplash.jpg}
%     \captionsetup{labelformat=empty}
%     \caption{\textbf{Credit}: M. Annamalai (Unsplash)}~\label{fig:marginfig}
%   \end{minipage}
% \end{marginfigure}

The connected scatterplot is not alone in its capacity to draw attention to the weird.
Visualization consultant Maarten Lambrechts has amassed a catalog of techniques that he calls \textbf{xenographics}~\cite{xenographics}, or \textsl{``weird but (sometimes) useful charts.''} 
The growing corpus of xenographics suggests that given a weird observation in data, there may (or should) exist a suitably weird way to represent it.

Beyond their capacity to herald the presence of weird patterns, xenographics are also worth discussing for another reason. 
Several documented xenographics could be described as hybrids or mash-ups of conventional chart types (e.g., a stacked area alluvial diagram, a scatterplot of line charts). 
A hybridization of form is also characteristic of \textbf{grotesque} art and architecture, in which the forms of animal and plant, or the organic and inorganic, are combined in ways that defy the conventions of representational art. 
According to Fisher, the grotesque is weird: grotesque works alternately fascinate and repel viewers, ultimately prompting a reaction that they seem \textsl{wrong}.
Hybrid xenographics could thus be thought of as a grotesque class of data visualization techniques, designed to communicate what no single chart could do on its own\footnote{Information graphics that meld figurative and abstract elements, such as Nigel Holmes~\cite{holmes2023}'s use of images as axes or coordinate spaces (i.e.,~\textsl{figurative frames}~\cite{byrne2019figurative}) could also be characterized as \textsl{grotesque}.}.

One final aspect of the weird worth considering here is the impression of not belonging, that the origin of the weird entity is some other domain.
Fisher discusses how this aspect of the weird is reinforced via visual imagery signifying an opening between the familiar and the unfamiliar: \textsl{``The centrality of doors, thresholds, and portals means that the notion of \textbf{the between} is crucial to the weird.''}
For me, this suggests an opportunity to explore these motifs when revealing weird data observations, with deliberate and metaphorical animations or interactive affordances: opening a door, parting a curtain, or traveling through a tunnel.

\section{Visualizing the Eerie}

Fisher's definition of the eerie as a \textsl{failure} \textsl{of} \textsl{presence} brought to mind Nicole Hengesbach's 2021 Information+ presentation~\cite{hengesbach2022seeing}, which included a review of the four types of \textbf{missingness} in data, referring to a case study on the mapping of urban greenery. 
If an entity (e.g., a tree) exists in reality and is either completely or partially captured as data, Hengesbach and colleagues would characterize the representation as \textsl{complete} (or at least partially complete).
However, if the entity is present in reality but unaccounted for in the data, this is an \textsl{absence}, while the converse is an \textsl{emptiness}. 
Lastly, there is \textsl{nothingness}: entities are neither present in reality or represented in the data.
Given these categories or missingness, an absence or an emptiness could be seen as eerie if there is ambiguity with respect to the processes of data collection and redaction: who was responsible, what their motivations were, and if any external forces acted to bring about absences or emptiness.
Nothingness can also be eerie, even when the processes of data collection are relatively transparent, and particularly in cases where \textsl{something} or \textsl{anything} was expected\footnote{e.g., visualizing the nothingness of space~\cite{worth} or ocean depths~\cite{agarwal}.}.

% \begin{marginfigure}
%   \begin{minipage}{\marginparwidth}
%     \centering
%     \includegraphics[width=\marginparwidth]{figures/Líneas_de_Nazca,_Nazca,_Perú,_2015-07-29,_DD_49.jpeg}
%     \captionsetup{labelformat=empty}
%     \caption{\textbf{Credit}: D. Delso (Wikimedia)}~\label{fig:marginfig}
%   \end{minipage}
% \end{marginfigure}

% \begin{marginfigure}
%   \begin{minipage}{\marginparwidth}
%     \centering
%     \includegraphics[width=\marginparwidth]{figures/CropCircleW.jpg}
%     \captionsetup{labelformat=empty}
%     \caption{\textbf{Credit}: Jabberocky (Pub. Domain)}~\label{fig:marginfig}
%   \end{minipage}
% \end{marginfigure}

% \begin{marginfigure}
%   \begin{minipage}{\marginparwidth}
%     \centering
%     \includegraphics[width=\marginparwidth]{figures/pelly-benassi-Hz1WQbHcXag-unsplash.jpg}
%     \captionsetup{labelformat=empty}
%     \caption{\textbf{Credit}: P. Benassi (Unsplash)}~\label{fig:marginfig}
%   \end{minipage}
% \end{marginfigure}

In communicative visualization design, evoking an eerie affect around absences in data requires establishing what expected patterns look like or how they have manifested in the past, contrasting presence with absence, and withholding or deferring possible explanations for this absence. 
An example of this formula is Isao Hashimoto's \textsl{1945 -- 1998}~\cite{hashimoto2012}, a multimodal combination of visualization and sonification, animating every atomic detonation during this time span on a world map, with different colors and tones indicating the nation responsible for each detonation. 
Following several cacophonous decades with hundreds of detonations and a barrage of color and sound, the last decade is markedly quieter: a relatively \textbf{eerie calm} has set in.
The YouTube video description includes the statement: \textsl{``No letter is used for equal messaging to all viewers without language barrier,''} however I maintain that by withholding any verbal or written explanation in the video itself accentuates an eerie affect. 
If it were possible to somehow contain work like this in a time capsule for a human (or non-human) civilization to discover in the distant future\footnote{e.g., information visualized on Voyager's Golden Record or on the moon-bound sapphire discs of the planned Sanctuary project~\cite{sanctuary}.}, I expect that a sense of the eerie would remain even if the viewer cannot readily interpret its meaning. 
The viewer may not only question what these audiovisual patterns represent, but also what or who is responsible for their cessation, similar how we currently speculate about Stonehenge, the Nazca Lines, crop circles, and other eerie traces of earlier / alien civilizations.

If a multimodal contrast between hectic noise and calm silence can elicit the eerie, \textbf{animation} design is another approach though which to reinforce a sense of the eerie in representations of data. If viewers are to believe that an unseen force or agent is acting upon what the data points represent, their disappearance could be animated in such a way that metaphorically suggests the presence of such forces, such as via simulated physics (e.g., a breeze blows the data points away) or via biomimicry~\cite{eggermont2018bio} (e.g., a school / flock of data points scatter and swim / fly away).

An eerie affect induced by a \textsl{failure of absence} could arguably be easily triggered given our propensity to \textbf{apophenia}, a bias in which we recognize patterns and attribute their appearance to an entity (or entities) that have agency. 
In most cases, however, there is no agent, and the pattern is fleeting, illusory, or insignificant. 
\textbf{Pareidolia} is a form of apophenia that is specific to visual stimuli, and visualization designers must exercise care to ensure that their viewers do not succumb to pareidolia unless this is the intent, such as in Michael Brenner's humorous \textsl{Viz} \textsl{in} \textsl{the} \textsl{Wild} project~\cite{brenner2022}, in which spurious visual patterns appearing in photographs are captioned as charts.
However, in select cases, there \textsl{is} an explanation for salient patterns where none were expected or designed. 
Dietmar Offenhuber's notion of \textbf{autographic} \textbf{visualization}~\cite{offenhuber2019data} is useful here, as the interpretation of material traces left by environmental phenomena\footnote{e.g., visual patterns of growth, oscillation, decay, oxidation, saturation, the accumulation of particulate matter, etc.} can often yield meaningful measurements as well as revelations regarding the process of data generation.
Meaningful measurements depend on several design operations, including the juxtaposition of frames, annotations, and decoding scales alongside material traces.
By undertaking these operations and measurements, practitioners can arrive at explanations for the patterns, which may include an attribution of agency, such as air pollution caused by human industry and transportation.
A sense of the eerie is instantiated when an autographic visualization interpretation either fails to attribute this agency or fails to explain the motivations of an identifiable agent.
Lastly, it is similarly eerie when an expected material trace vanishes, and we are bereft of any explanation for this disappearance.
I'll offer a personal example: for years I regularly visited a Florida beach that was pockmarked with tiny holes made by sand fleas, a visual pattern distributed evenly along the shoreline.
One year, this pattern had vanished altogether along one stretch of the beach, and in that moment I could not explain this lifeless failure of presence\footnote{I would later learn that the adjacent resort hotel had began importing sand from elsewhere, destroying the natural habitat of the fleas.}. 

% \begin{marginfigure}
%   \begin{minipage}{\marginparwidth}
%     \centering
%     \includegraphics[width=\marginparwidth]{figures/West_Palm_Beach_high_rises.jpg}
%     \captionsetup{labelformat=empty}
%     \caption{\textbf{Credit}: K. Pardi (Wikimedia)}~\label{fig:marginfig}
%   \end{minipage}
% \end{marginfigure}

\section{Implications for Visualization Design}

Until this point, my essay has been predominantly reflective, an attempt at connecting the visualization community with ideas that originated from a critical analysis of popular culture. 
However, I will nevertheless try to summarize some prescriptive implications for visualization design, or at least suggest some creative experiments to undertake:

% \begin{marginfigure}
%   \begin{minipage}{\marginparwidth}
%     \centering
%     \includegraphics[width=\marginparwidth]{figures/patty-zavala-VWJZ5v9-SZ0-unsplash.jpg}
%     \captionsetup{labelformat=empty}
%     \caption{\textbf{Credit}: P. Zavala (Unsplash)}~\label{fig:marginfig}
%   \end{minipage}
% \end{marginfigure}

% \begin{marginfigure}
%   \begin{minipage}{\marginparwidth}
%     \centering
%     \includegraphics[width=\marginparwidth]{figures/20181204_Warming_stripes_(global,_WMO,_1850-2018)_-_Climate_Lab_Book_(Ed_Hawkins).png}
%     \captionsetup{labelformat=empty}
%     \caption{\textbf{Credit}: E. Hawkins (Wikimedia)}~\label{fig:marginfig}
%   \end{minipage}
% \end{marginfigure}

\setlist{nolistsep}
\begin{itemize}[leftmargin=*,noitemsep]
    \item Acknowledge situations where it is appropriate for data visualization to make viewers uncomfortable.
    \item Employ egocentric perspectives and doorway motifs for distinguishing the familiar from the weird.
    \item Avoid manipulating viewers' impressions by pairing an expected or familiar data observation with a weird choice of representation, reserving the grotesque art of xenographics for communicating truly weird patterns in data.
    \item Withhold or delay revelatory explanations or annotations for failures of absence and failures of presence.
    \item Provide subtle indications of agency via animation.
\end{itemize}

\section{Visualization for Weird and Eerie Times}

We are living in \textsl{weird} times~\cite{permaweird}. 
Around the world, we have experienced weird weather fluctuations in recent years.
Where I live now, the color of the sky can be a weird dull orange in late summer.
Meanwhile, invasive species and algal blooms have altered ecosystems.
While I was growing up, there did not used to be ticks where I lived, but these weirdly resilient insects and the threat of Lyme disease can now be found there.
In the economic sphere, sequences of weird events continue to disrupt currencies, real estate markets, and supply chains. 
How we represent these and other weird happenings, whenever we capture them as data, should be commensurately weird, particularly if we are to communicate just how weird they are.
For example, conventions in meteorological visualization may not capture how weird these extreme weather events are. 
New visual idioms like Ed Hawkins' \textit{Warming Stripes}~\cite{Hawkins2018} are useful in this regard, though even these elicit questions of how we should visualize future extreme values: while we could  renormalize the color palettes to make deep blues and deep reds more apparent, renormalization may fail to capture the severity of the next weird event. 

Our world is also increasingly eerie.
Jenny Odell recently wrote about the feeling of living in \textsl{apocalyptic} \textsl{times}~\cite{odell2023savingtime}, which is fitting given how post-apocalyptic films and novels often evoke a sense of the eerie. 
Biodiversity loss has resulted in eerie landscapes and oceans: why are so many species absent? 
Human migration has resulted in eerily empty urban centers, empty rural settlements, and conversely crowded interstitial spaces: why are people absent in places where they are expected, but present in places where they are not? 
Odell's writing also mentions that the etymology of \textsl{apocalypse} is Greek, meaning \textsl{to reveal}, and that prior to its modern English usage signifying `an ending', apocalypse was closer in meaning to \textsl{insight}.
Given the old adage \textsl{``the purpose of visualization is insight, not pictures''}~\cite{card1999readings}, could there be some utility in the practice of \textsl{apocalyptic data visualization}?
By visualizing the data from the apocalypses unfolding around us, and withholding our attempts to explain the absences and presences therein, we have the power to invoke the eerie. 
From this eerie experience, each viewer can arrive at some degree of insight, that is, to identify the forces responsible. 

For much of what could be constituted as weird or eerie in the capitalocene, we could (like Fisher) attribute agency to movements of capital, to movements of an invisible hand.
Could we more specific than this?
With data visualization, we have the potential to lend these attributions a semblance of greater precision, and ultimately, the potential to change what viewers believe.

\section{Acknowledgments}

Thanks to M. Correll, A. McNutt, and the alt.VIS reviewers.

\balance{} 

\bibliographystyle{SIGCHI-Reference-Format}
\bibliography{extended-abstract}


\begin{thebibliography}{00}

%%% the Web address.

\ifx \showCODEN    \undefined \def \showCODEN     #1{\unskip}     \fi
\ifx \showDOI      \undefined \def \showDOI       #1{{\tt DOI:}\penalty0{#1}\ }
  \fi
\ifx \showISBNx    \undefined \def \showISBNx     #1{\unskip}     \fi
\ifx \showISBNxiii \undefined \def \showISBNxiii  #1{\unskip}     \fi
\ifx \showISSN     \undefined \def \showISSN      #1{\unskip}     \fi
\ifx \showLCCN     \undefined \def \showLCCN      #1{\unskip}     \fi
\ifx \shownote     \undefined \def \shownote      #1{#1}          \fi
\ifx \showarticletitle \undefined \def \showarticletitle #1{#1}   \fi
\ifx \showURL      \undefined \def \showURL       #1{#1}          \fi

\bibitem{agarwal}
{Neal Agarwal}. n.d.
\newblock The Deep Sea.
\newblock \href{https://neal.fun/deep-sea/}{neal.fun/deep-sea}.   (n.d.).
\newblock


\bibitem{bertini2020shouldn}
{Enrico Bertini}, {Michael Correll}, {and} {Steven Franconeri}. 2020.
\newblock \showarticletitle{Why shouldn't all charts be scatter plots? Beyond
  precision-driven visualizations}. In {\em Short paper proceeedings of the
  IEEE Visualization Conference (VIS)}.
\newblock
\showDOI{%
\url{https://dx.doi.org/10.1109/VIS47514.2020.00048}}


\bibitem{brenner2022}
{Michael Brenner}. 2022.
\newblock \showarticletitle{Viz in the wild}.
\newblock
  \href{https://www.instagram.com/vizinthewild/}{instagram.com/vizinthewild},
  {\em Nightingale\/}  {1} (2022).
\newblock


\bibitem{byrne2019figurative}
{Lydia Byrne}, {Daniel Angus}, {and} {Janet Wiles}. 2019.
\newblock \showarticletitle{Figurative frames: A critical vocabulary for images
  in information visualization}.
\newblock {\em Information Visualization\/} {18}, 1 (2019).
\newblock
\showDOI{%
\url{https://dx.doi.org/10.1177/1473871617724212}}


\bibitem{card1999readings}
{Stuart~K Card}, {Jock Mackinlay}, {and} {Ben Shneiderman}. 1999.
\newblock {\em Readings in Information Visualization: Using Vision to Think}.
\newblock Morgan Kaufmann.
\newblock


\bibitem{eggermont2018bio}
{Marjan~Jos{\'e} Eggermont}. 2018.
\newblock {\em Bio-inspired Design and Information Visualization}.
\newblock Ph.D. Dissertation.
\newblock
\newblock
\shownote{University of Calgary.}


\bibitem{fisher2016}
{Mark Fisher}. 2016.
\newblock {\em The Weird and the Eerie}.
\newblock Repeater Books.
\newblock


\bibitem{gaver2003ambiguity}
{William~W Gaver}, {Jacob Beaver}, {and} {Steve Benford}. 2003.
\newblock \showarticletitle{Ambiguity as a resource for design}. In {\em
  Proceedings of the ACM Conference on Human Factors in Computing Systems
  (CHI)}.
\newblock
\showDOI{%
\url{https://dx.doi.org/10.1145/642611.642653}}


\bibitem{Halloran}
{Neil Halloran}. 2016---.
\newblock Data-driven documentaries.
\newblock
  \href{https://www.youtube.com/@NeilHalloran}{youtube.com/@NeilHalloran}.
  (2016---).
\newblock


\bibitem{haroz2015connected}
{Steve Haroz}, {Robert Kosara}, {and} {Steven~L Franconeri}. 2015.
\newblock \showarticletitle{The connected scatterplot for presenting paired
  time series}.
\newblock {\em IEEE Transactions on Visualization and Computer Graphics
  (TVCG)\/} {22}, 9 (2015).
\newblock
\showDOI{%
\url{https://dx.doi.org/10.1109/TVCG.2015.2502587}}


\bibitem{hashimoto2012}
{Isao Hashimoto}. 2012.
\newblock 1945---1998.
\newblock \href{https://youtu.be/cjAqR1zICA0}{youtu.be/cjAqR1zICA0}.   (2012).
\newblock


\bibitem{Hawkins2018}
{Ed Hawinks}. 2018.
\newblock Warming stripes.
\newblock
  \href{https://www.climate-lab-book.ac.uk/2018/warming-stripes/}{climate-lab-book.ac.uk/2018/warming-stripes/}.
    (2018).
\newblock


\bibitem{hengesbach2022seeing}
{Nicole Hengesbach}, {Greg~J McInerny}, {and} {Jo{\~a}o Porto~de Albuquerque}.
  2022.
\newblock \showarticletitle{Seeing what is not shown: Combining visualization
  critique and design to surface the limitations in data}.
\newblock {\em Information Design Journal\/} {27}, 1 (2022).
\newblock
\showDOI{%
\url{https://dx.doi.org/10.1075/idj.22006.hen}}


\bibitem{holmes2023}
{Nigel Holmes}. 2023.
\newblock {\em Joyful Infographics: A Friendly, Human Approach to Data}.
\newblock CRC Press.
\newblock


\bibitem{hullman2013contextifier}
{Jessica Hullman}, {Nicholas Diakopoulos}, {and} {Eytan Adar}. 2013.
\newblock \showarticletitle{Contextifier: automatic generation of annotated
  stock visualizations}. In {\em Proceedings of the ACM Conference on Human
  Factors in Computing Systems (CHI)}.
\newblock
\showDOI{%
\url{https://dx.doi.org/10.1145/2470654.2481374}}


\bibitem{KennyBecker}
{Roger Kenny} {and} {{Ana Asnes} Becker}. 2015.
\newblock Is the Nasdaq in Another Bubble? A virtual reality guided tour of 21
  years of the Nasdaq.
\newblock
  \href{http://graphics.wsj.com/3d-nasdaq/}{graphics.wsj.com/3d-nasdaq}.
  (2015).
\newblock


\bibitem{Kirk2014}
{Andy Kirk}. 2014.
\newblock The Design of Nothing: Null, Zero, Blank.
\newblock OpenVisConf 2014 Conference presentation:
  \href{https://youtu.be/JqzAuqNPYVM}{youtu.be/JqzAuqNPYVM}.   (2014).
\newblock


\bibitem{kosara2016presentation}
{Robert Kosara}. 2016.
\newblock \showarticletitle{Presentation-oriented visualization techniques}.
\newblock {\em IEEE Computer Graphics and Applications (CG\&A)\/} {36}, 1
  (2016).
\newblock
\showDOI{%
\url{https://dx.doi.org/10.1109/MCG.2016.2}}


\bibitem{xenographics}
{Maarten Lambrechts}. 2018---.
\newblock Xenographics: Weird but (sometimes) useful charts.
\newblock \href{https://xeno.graphics/}{https://xeno.graphics}.   (2018---).
\newblock


\bibitem{lee2022affective}
{Elsie Lee-Robbins} {and} {Eytan Adar}. 2023.
\newblock \showarticletitle{Affective learning objectives for communicative
  visualizations}.
\newblock {\em IEEE Transactions on Visualization and Computer Graphics
  (Proceedings of VIS)\/} {29}, 1 (2023).
\newblock
\showDOI{%
\url{https://dx.doi.org/10.1109/TVCG.2022.3209500}}


\bibitem{marshall2011gas}
{Joe Marshall}, {Brendan Walker}, {Steve Benford}, {George Tomlinson}, {Stefan
  Rennick~Egglestone}, {Stuart Reeves}, {Patrick Brundell}, {Paul Tennent}, {Jo
  Cranwell}, {Paul Harter}, {and} {others}. 2011.
\newblock \showarticletitle{The gas mask: a probe for exploring fearsome
  interactions}.
\newblock In {\em Extended Abstract Proceedings of the ACM Conference on Human
  Factors in Computing Systems (alt.CHI)}.
\newblock
\showDOI{%
\url{https://dx.doi.org/10.1145/1979742.1979609}}


\bibitem{xkcd}
{Randall Munroe}. 2005---.
\newblock {xkcd}.
\newblock \href{https://xkcd.com/}{xkcd.com}.   (2005---).
\newblock


\bibitem{odell2023savingtime}
{Jenny Odell}. 2023.
\newblock {\em Saving Time: Discovering a Life Beyond the Clock}.
\newblock Random House.
\newblock


\bibitem{offenhuber2019data}
{Dietmar Offenhuber}. 2020.
\newblock \showarticletitle{Data by proxy --- material traces as autographic
  visualizations}.
\newblock {\em IEEE Transactions on Visualization and Computer Graphics
  (Proceedings of InfoVis)\/} {26}, 1 (2020), 98--108.
\newblock
\showDOI{%
\url{https://dx.doi.org/10.1109/TVCG.2019.2934788}}


\bibitem{pandey2014persuasive}
{Anshul~Vikram Pandey}, {Anjali Manivannan}, {Oded Nov}, {Margaret
  Satterthwaite}, {and} {Enrico Bertini}. 2014.
\newblock \showarticletitle{The persuasive power of data visualization}.
\newblock {\em IEEE Transactions on Visualization and Computer Graphics
  (Proceedings of InfoVis)\/} {20}, 12 (2014).
\newblock
\showDOI{%
\url{https://dx.doi.org/10.1109/TVCG.2014.2346419}}


\bibitem{permaweird}
{Venkatesh Rao}, {Holly Herndon}, {and} {Mat Dryhurst}. 2023.
\newblock {The Permaweird with Venkatesh Rao}.
\newblock \textsl{Interdependence} episode 95 (April 27, 2023),
  \href{https://interdependence.fm/episodes/the-permaweird-with-venkatesh-rao}{interdependence.fm}.
    (2023).
\newblock


\bibitem{sanctuary}
Sanctuary Project n.d.
\newblock {Sanctuary Project}.
\newblock \href{https://www.sanctuaryproject.eu/}{sanctuaryproject.eu}.
  (n.d.).
\newblock


\bibitem{vigen2015}
{Tylen Vigen}. 2015.
\newblock {\em Spurious Correlations}.
\newblock Hachette Books.
\newblock


\bibitem{wilkinson2006high}
{Leland Wilkinson}, {Anushka Anand}, {and} {Robert Grossman}. 2006.
\newblock \showarticletitle{High-dimensional visual analytics: Interactive
  exploration guided by pairwise views of point distributions}.
\newblock {\em IEEE Transactions on Visualization and Computer Graphics
  (TVCG)\/} {12}, 6 (2006).
\newblock
\showDOI{%
\url{https://dx.doi.org/10.1109/TVCG.2006.94}}


\bibitem{worth}
{Josh Worth}. n.d.
\newblock If the Moon were Only 1 Pixel.
\newblock
  \href{https://joshworth.com/dev/pixelspace/pixelspace_solarsystem.html}{joshworth.com}.
    (n.d.).
\newblock


\end{thebibliography}

\end{document}